\documentclass[usenatbib]{mn2e}

\newif\ifpdf
\ifx\pdfoutput\undefined
\pdffalse 
\else
\pdfoutput=1 
\pdftrue
\fi

\ifpdf
\usepackage[pdftex]{graphicx}
\DeclareGraphicsExtensions{.pdf, .jpg}
\else
\usepackage[dvips]{graphicx}
\DeclareGraphicsExtensions{.eps, .jpg}
\fi

\usepackage[boxed]{algorithm}
\usepackage{algorithmic}

\def\dom{\mathcal{D}}
\def\diag{\mathrm{diag}}
\def\trace{\mathrm{tr}}
\def\inv{^{-1}}
\def\adj{^\dagger }
\def\invsq{^{-\frac 1 2 }}
\def\degree{^\circ} 
\newcommand\mypar[1]{\medskip\par\noindent\textbf{#1}\ \ }

\title[MDMC spectral matching for CMB data analysis]
{Multi--Detector Multi--Component spectral matching and
applications for CMB data analysis}
\author[J. Delabrouille, J.-F. Cardoso, G. Patanchon]{ J. Delabrouille$^1$, J.-F. Cardoso$^2$, G. Patanchon$^1$ \\
$^1$ PCC --- Coll{\`e}ge de France, 11, place Marcelin Berthelot, F-75231 Paris, France\\
$^2$ CNRS/ENST --- 46, rue Barrault, 75634 Paris, France
}
\date{Accepted 2002 January 00.
      Received 2002 January 00;
      in original form 2002 January 00}
\pagerange{\pageref{firstpage}--\pageref{lastpage}}
\pubyear{2002}
\begin{document}

\maketitle
\label{firstpage}

\begin{abstract}
  We present a new method for analyzing multi--detector maps
  containing contributions from several components. Our method, based
  on matching the data to a model in the spectral domain, permits to
  estimate jointly the spatial power spectra of the components and of
  the noise, as well as the mixing coefficients. It is of particular
  relevance for the analysis of millimeter--wave maps containing a
  contribution from CMB anisotropies.
\end{abstract}


\section{Introduction}

Mapping sky emissions at millimeter wavelengths, and in particular
Cosmic Microwave Background (CMB) anisotropies, is one of the main
objectives of ongoing observational effort in millimeter-wave
astronomy.  Sensitive balloon--borne and space--borne missions such as
Archeops \citep{2002APh....17..101B}, 
Boomerang \citep{2000Natur.404..955D},
Maxima \citep{2000ApJ...545L...5H} 
and MAP \citep{1997AAS...191.8701B}
are currently in operating status, yielding a large amount of
multi--detector and multi--frequency measurements.  Within a few
years, the Planck mission \citep{lamarre00,bersanelli00}, to be
launched by ESA in 2007, will observe the complete sky with $\sim 100$
detectors distributed in nine frequency bands ranging from 30 to 850
GHz.  The main objective of these observations is the determination of
the spatial power spectrum of CMB anisotropies. A secondary objective
is identifying and mapping the emission from all contributing
astrophysical processes.

The availability of several detectors operating in several bands makes
it possible to devise new powerful data processing schemes.  In
particular, by combining data from several detectors, it is possible
to improve substantially the signal-to-noise ratio (by weighted
averaging) and to separate several foreground components (possibly of
astrophysical interest in their own right) from the CMB by component
separation methods.
Component separation, however, typically requires a good knowledge of 
the transfer function connecting a multi-component sky to multi-detector maps. 

This paper proposes to use spectral matching as a new approach to
processing multi-detector multi-component (MDMC) data, in which all
the information needed to estimate the spatial power spectra of
components \emph{and/or} to separate them is sought in the data
structure itself.  The method works with or without prior detector
calibration and gives access to spatial power spectra in a
straightforward way; it is statistically efficient (being a maximum
likelihood technique) and computationally efficient (working with a
small set of sufficient statistics rather than with original maps).

This paper is organised as follows.  The idea of spectral estimation
via multi--detector multi--component spectral matching is introduced
in section~\ref{sec:spectr-match-appr}.
Section~\ref{sec:blind-separ-meth} describes the technique in more
detail, connects it to a maximum likelihood method, and discusses
the specific implementations.
Section~\ref{sec:testing} is devoted to evaluating the performance of
the method on synthetic Planck HFI observations.
We discuss the method and extensions in section~\ref{sec:discussion}.

\section{The multi-detector multi-component framework}\label{sec:spectr-match-appr}

Multi-detector CMB measurements can be modeled as resulting from the
superposition of multiple components. Statistically efficient data
processing should coherently exploit this MDMC structure.

\bigskip

The sky emission at millimeter wavelengths is well modeled at first
order by a linear superposition of the emissions of a few processes:
CMB anisotropies, thermal dust emission, thermal Sunyaev Zel'dovich
(SZ) effect, synchrotron emission, etc.  The observation of the sky
with detector $d$ is then a noisy linear mixture of $N_c$
components:
\begin{equation}
  y_{d}(\theta, \phi) 
  = 
  \sum_{j=1}^{N_c} A_{d j} s_{j}(\theta, \phi) + 
  n_{d}(\theta, \phi)
  \label{eq:mixture}
\end{equation}
where $s_{j}$ is the emission template for the $j$th astrophysical
process, herein referred to as a \emph{source} or a \emph{component}.
The coefficients $A_{dj}$ reflect emission laws and detector
properties while $n_d$ accounts for noise. For simplicity, we neglect for the moment beam effects, postponing the discussion to section~\ref{sec:discussion}.

Quantities of prime interest are spatial power spectra.  For the
$j$-{th} component, at frequency $\vec \ell$, this is:
\begin{equation}
  C_j(\vec \ell) = \langle |s_j(\vec \ell)|^2 \rangle
\end{equation}
where $\langle\cdot\rangle$ denotes the expectation operator and $\vec
\ell$ indexes either a Fourier mode or an $(\ell,m)$ mode.

In practice, power spectra are estimated by averages over bins:
\begin{equation}
    C_j(q) = \frac 1 {n_q} \sum_{\vec\ell\in\dom_q} C_j(\vec \ell) 
    .
\label{eq:Cjq}
\end{equation}
where $q=1,\ldots,Q$ is the spectral bin index, $\dom_q$ is the set of
frequencies contributing to bin $q$ and $n_q$ is the number of such
frequencies.\footnote{It is customary for CMB data analysis to weight the terms in sum~\ref{eq:Cjq} by $\ell(\ell+1)$. For the sake of exposition, we use a flat weighting here (see section~\ref{sec:discussion} for weighted sums)} Typical bins can be bands $\ell_{\rm min} \leq \ell < \ell_{\rm max}$ extending over a range of one to tens of $\ell$ values.

\mypar{Multi-detector power spectrum}

Since we focus on jointly processing the maps from all detectors, it
is convenient to stack $y_1, \ldots, y_{N_d}$ into a single $N_d\times
1$ vector $Y$.
Then, the set of eqs.~\ref{eq:mixture} for all $N_d$ detectors is more compactly
written in matrix--vector form as:
\begin{equation}\label{eq:Mixture}
  Y(\theta, \phi) = A S(\theta, \phi) + N(\theta, \phi)
\end{equation}
with a so called $N_d\times N_c$ `mixing matrix' $A$.
In Fourier space, this equation reads
\begin{equation}
  Y(\vec \ell) = A S(\vec \ell) + N(\vec \ell) .
\end{equation}
The power spectrum of process $Y$ is represented by the $N_d\times
N_d$ spectral density matrix $\langle Y(\vec\ell) Y(\vec\ell)\adj
\rangle$ where $\cdot\adj$ denotes transpose-conjugation.
Its average over bins
\begin{equation}
  R_Y(q) = \frac 1 {n_q} \sum_{\vec\ell\in\dom_q}  \langle  Y(\vec\ell) Y(\vec\ell)\adj \rangle
  \ \ \ \ (q=1,\ldots, Q)
\end{equation}
will also be referred to as a spectral density matrix.  
According to the linear model~(\ref{eq:mixture}), it is structured as:
\begin{equation}
  \label{e:strucSCM}
  R_Y(q) = A R_S(q) A\adj + R_N(q)
  \ \ \ \ (q=1,\ldots, Q)
\end{equation}
with $R_S(q)$ and $R_N(q)$ defined similarly to $R_Y(q)$.
Statistical independence between components implies:
\begin{equation}
  \label{e:structRs}
  R_S(q) 
  =
  \diag\left( C_1(q), \ldots, C_{N_c}(q) \right)
  .
\end{equation}
For the sake of exposition, we assume that the noise is uncorrelated,
both across detectors and in space, so that the noise structure is
described by $N_d$ parameters:
\begin{equation}
  \label{e:Rnq}
  R_N(q) = \diag\left(\sigma_1^2, \ldots, \sigma_{N_d}^2\right)
  .  
\end{equation}

\mypar{Parameter extraction by spectral matching}

The MDMC model, as defined by
eqs.~(\ref{e:strucSCM}-\ref{e:structRs}-\ref{e:Rnq}), depends on a set
$\{ R_Y(q) \}$ of $Q$ spectral density matrices, which in turn depend
on $\{ A, C_j(q), \sigma_d^2) \}$, amounting to $N_d\times N_c+
Q\times N_c + N_d$ scalar parameters.
However, the number of independent correlations in $Q$ spectral
density matrices is $Q\times N_d(N_d+1)/2$ (since each matrix is real
symmetric).
This later number is (in general) higher than the former.

With this in mind, our proposal can be summarized as `MDMC spectral
matching', meaning: \emph{estimate all (or parts of) the parameters
$\{ A, C_j(q), \sigma_d^2) \}$ by finding the best match between $\{
R_Y(q) \}$, as specified
by~(\ref{e:strucSCM}-\ref{e:structRs}-\ref{e:Rnq}), and a set of $Q$
`empirical spectral density matrices' $\{ \widehat R_Y(q)\}$}:
\begin{equation}
  \label{e:defspeccovmat}
  \widehat R_Y(q) 
  =
  \frac 1 {n_q} \sum_{\vec\ell\in\dom_q}  Y(\vec\ell) Y(\vec\ell)\adj
  \ \ \ \ (q=1,\ldots, Q)
\end{equation}
which are the natural non parametric estimates of the corresponding
$R_Y(q)$.

Some preliminary comments about the MDMC spectral matching approach
are in order.
  
\mypar{Parameter choice:}
  There is a lot of flexibility in the choice of parameters over which
  to minimize the spectral mismatch.  By selecting different sets of
  parameters, different goals can be achieved.
  For instance, we may assume that matrix $A$ and the noise spectrum
  $R_N(q)$ are known so that the mismatch is minimized only with
  respect to the binned spectra $C_i(q)$ of all components: the method
  appears as a spectral estimation technique \emph{which does not
  require the explicit separation of the observed maps into component 
  maps}.
  Another important example, as illustrated in section~\ref{sec:testing},
  consists in including matrix $A$ among the free parameters.  Then, the
  method works as the so-called `blind techniques', and permits the 
  measurement of the emission law of the components, or the cross 
  calibration of detectors.

\mypar{Degeneracies:}
  A key issue in spectral matching is whether or not matrix $A$ can be
  uniquely determined from the data only.  When all parameters $\{ A,
  C_j(q), \sigma_d^2 \}$ are allowed to be adjusted, there are at least
  two clear indeterminations. 
  First, the \emph{ordering} (or numbering) of the components in the
  model is immaterial: matrix $A$ cannot be recovered better than up
  to column permutation on the sole basis of a spectral match.
  Second, a scalar factor can be exchanged, for each component $j$,
  between the $j$th column of $A$ and $C_j(q)$.  These scale factors
  cannot be determined from the data themselves.  
  
  Another trivial case of indetermination is when two columns of $A$ 
  corresponding to physically distinct components are proportional. In 
  this case, the sum of the two appears in the model as one single 
  component. The identifiability of the other components is not affected.
  
  A more severe degeneracy occurs if any two components have
  proportional spectra.  In this case, as is known from the noiseless
  case~\citep{Pham:IEEESP:97}, only the space spanned by the
  corresponding columns of $A$ can be determined in a spectral match
  with $A$ as a free parameter.  In this case however, the
  identifiability of the other components is unchanged, with no impact
  on the accuracy of component separation with a Wiener method
  (sec.~\ref{sec:discussion}).  The key point to remember is that
  spectral matching requires spectral diversity to separate components
  associated with unknown columns of $A$.

\mypar{Maximum likelihood:}
  Section~\ref{sec:spectral-mismatch} explains why `spectral matching'
  corresponds to maximum likelihood estimation.
  This happens in a Gaussian stationary model with smooth (actually:
  constant over bins) spectra.  In such a model the likelihood of the
  observations \emph{is} a measure~(\ref{eq:obj}) of spectral
  matching.  Since the likelihood then depends on the data \emph{only}
  via the empirical spectral density matrices, the massive data
  reduction gained from replacing the observations by a (usually) much
  smaller set of statistics (the empirical spectral density matrices
  $\widehat R_Y(q)$) is obtained without information loss.

\mypar{Comparison with component separation:}
It is interesting to compare spectral matching to techniques based on
prior explicit component separation.

Producing a CMB map as free as possible from foreground and noise
contamination is the objective of the component separation step, in
which maps obtained at different frequencies are combined to maximize
the signal to noise ratio (where noise includes also foreground
contamination).

  The usual approach for taking advantage of multi--detector
  measurements can be summarised as:
  first, form estimates $\widehat s_j(\vec \ell)$ of component maps 
  $s_j(\vec \ell)$ (via component separation), 
  second, estimate the spectrum of each component $j$ by averaging 
  within bins:
  \begin{equation}
    \widehat C_j(q)
    =
    \frac 1 {n_q} \sum_{\vec\ell\in\dom_q} | \widehat s_j(\vec \ell) |^2
  \end{equation}
  with, possibly, some post-processing of the power spectrum estimates.

This method suffers from two difficulties. First, the best component 
separation methods typically require the prior knowledge of the 
statistical properties of the components (including the CMB power 
spectrum) and of the noise. Second, recovered maps contain residuals 
(including noise) which contribute to the total power, biasing the 
spectrum estimated on the map, unless the power spectrum of these 
residuals can be estimated accurately and subtracted for de--biasing.

In contrast our approach takes the reverse path. The first step is the 
estimation of the spectrum for the multi--detector map (which takes the 
form of a sequence of spectral density matrices). This first step 
preserves all the joint correlation structure between maps. In essence, 
the second step (spectral matching) amounts to resolving the 
joint power spectrum into spectra of individual components.

Hence, instead of first separating component \emph{maps} and then computing power spectra, we first compute the multivariate power spectrum and then separate component \emph{spectra}.

\section{MDMC spectral matching in practice}\label{sec:blind-separ-meth}

The implementation of MDMC spectral matching is now described in more detail.
Section~\ref{sec:spectral-mismatch} introduces the spectral matching
criterion;
section~\ref{sec:em} describes the EM algorithm for its optimization;
section~\ref{sec:nlin-optim} describes a complementary technique for fast convergence.

\subsection{Maximum likelihood spectral matching}\label{sec:spectral-mismatch}

Any reasonable measure of mismatch between the empirical density 
matrices $\{ \widehat R_Y (q) \}$ and their model counterparts
$\{\widehat R_Y (q; \theta) \}$ could be used to compute estimates of
a $\theta$ parameter.  
In order to get good estimates, however, one should use a mismatch
criterion derived from statistical principles.
Such a derivation can be based on the statistical distribution of the
Fourier coefficients of a stationary process which are (at least
asymptotically in the data size) normally distributed, uncorrelated,
with a variance proportional to the power spectrum (Whittle approximation, see appendix~\ref{sec:whittle}).
Thus, the likelihood of the observations can be readily expressed in
terms of spectral density matrices.
Appendix~\ref{sec:whittle} outlines how the (negative)
log-likelihood of the data then is (up to irrelevant factors and
terms) equal to
\begin{equation}\label{eq:obj}
  \phi(\theta)
  =
  \sum_{q=1}^Q
  n_q
  \
  D\left(\widehat R_Y (q),  R_Y(q; \theta) \right) 
\end{equation}
where $D(\cdot,\cdot)$ is a measure of divergence between two positive
$n\times n$ matrices defined by
\begin{equation}
  \label{eq:kullR}
  D(R_1,R_2) 
  =
  \trace \left( R_1R_2\inv \right) - \log\det (R_1R_2\inv) - n  
  .
\end{equation}
It can be seen\footnote{For instance by expressing $D(R_1,R_2)$ in
terms of the eigenvalues of $R_2\invsq R_1 R_2\invsq$.}  that
$D(R_1,R_2)\geq 0$ with equality if and only if $R_1=R_2$.
Thus \emph{spectral matching corresponds to maximum likelihood
estimation in a stationary model}. The minimizer of $\phi(\theta)$
is then a maximum likelihood estimate, and inherits the good statistical properties associated to it.

Only in an asymptotic framework can maximum likelihood procedures be
proved to reach minimum estimation variance.  It means that criteria
which are equivalent to~(\ref{eq:obj}) are expected to have the same
statistical quality as~(\ref{eq:obj}).
In particular, criterion~(\ref{eq:obj}) can be replaced by a quadratic
approximation: when each $\widehat R_Y(q)$ is close $R_Y(q;\theta)$, a
second-order expansion of $D(\widehat R_Y , R_Y)$ yields
\begin{equation}
  \label{e:phiquad}
  D_2\left(\widehat R_Y ,  R_Y \right) 
  =
  \trace
  \left(
  \widehat R_Y\inv 
  ( \widehat R_Y - R_Y )
  \widehat R_Y \inv
  ( \widehat R_Y  - R_Y )
  \right) 
  .
\end{equation}
The resulting quadratic criterion is of particular interest when the
unknown parameters enter \emph{linearly} in $R_Y(q;\theta)$
(for instance when $A$ is known and $\theta$ only contains the binned
power spectra of the components) since then criterion minimization
becomes trivial.
In this paper, however, we stick to
using~(\ref{eq:obj}-\ref{eq:kullR}).
Even though the divergence~(\ref{eq:kullR}) may, in the general case,
seem more difficult to deal with than its quadratic
approximation~(\ref{e:phiquad}), it actually lends itself to simple
optimization via the EM algorithm (see section~\ref{sec:em}) thanks to
its connection to the likelihood.

\subsection{The EM algorithm}\label{sec:em}

The expectation-maximization (EM) algorithm~\citep{Dempster77} is a
well known technique for maximizing the likelihood of statistical
models which include `latent' or `unobserved' variables.
It is well suited to our purpose by taking the components as the
latent variables.
The EM algorithm is iterative: starting from an initial value of the
parameters, it performs a sequence of parameter updates called
`EM-steps'.  Each step is guaranteed to increase the likelihood of the
parameters.

The spectral matching criterion~(\ref{eq:obj}) actually being a
likelihood function in disguise, the EM algorithm can be used for its
minimization.  Each EM step is guaranteed to improve the spectral fit
by decreasing $\phi(\theta)$.

We consider the regular EM algorithm, based on the Gaussian likelihood
described in appendix~\ref{sec:whittle} and taking as `latent
variables' the spectral modes $Y(\vec\ell)$.  The
form of the EM steps immediately follows as sketched in appendix~\ref{sec:domain-em} and summarized by the pseudo-code.

\begin{algorithm}
  \caption{The EM algorithm for minimizing the MDMC
  spectral mismatch $\phi(\theta)$ with respect to  
  $\theta= \{ A,C_j(q),\sigma_d^2 \}$.
  }
  \label{algo:1}
  \begin{algorithmic}
    \REQUIRE{Spectral density matrices $\widehat
    R_Y(1),\ldots,\widehat R_Y(Q)$} \REQUIRE{Initial value of $\theta
    =\left \{ A, C_j(q), \sigma_d^2 \right \}$. }  \STATE Set $\widetilde
    R_{yy}(q) = \widehat R_Y(q) $ and $\widetilde
    R_{yy}=\sum_{q}\frac{n_q}n \ R_{yy}(q)$.  \REPEAT \STATE \COMMENT
    {\hrulefill\ E-step.  Compute conditional statistics:} \STATE Set
    $R_S(q)=\diag(C_j(q))$ and $R_N=\diag(\sigma_d^2)$ \FOR {$q=1$ to
    $Q$} \STATE $G(q) = (A\adj R_N \inv A + R_S(q)\inv )\inv $ \STATE
    $W(q) = G(q) A\adj R_N\inv $ \STATE $\widetilde R_{ss}(q) = W(q)
    \widehat R_Y(q) W(q)\adj + G(q) $ \STATE $\widetilde R_{sy}(q) = W(q)
    \widehat R_Y(q) $ \ENDFOR \STATE $\widetilde R_{ss}=\sum_{q=1}^Q
    \frac{n_q}n \ \widetilde R_{ss}(q) $ \STATE $\widetilde
    R_{ys}=\sum_{q=1}^Q \frac{n_q}n \ \widetilde R_{ys}(q) $ \STATE
    \COMMENT {\hrulefill\ M-step.  Update the parameters:} \STATE $A =
    \widetilde R_{ys} \widetilde R_{ss}\inv$
    \STATE $C_i(q) =  \left[\widetilde  R_{ss}(q) \right]_{ii}$
    \STATE $\sigma_d^2 = \left[ \widetilde R_{yy} - \widetilde R_{sy}\adj  \widetilde R_{ss}\inv \widetilde  R_{sy} \right]_{dd} $
    \STATE Rescale the parameters (see text).
    \UNTIL {a convergence criterion is satisfied}
  \end{algorithmic}
\end{algorithm}
It is worth mentionning that EM steps take such a regular structure
when the parameters are $\theta=\{ A, C_j(q), \sigma_d^2 \}$.  A slightly
different form would result from a more constrained parameter set.

Recall that, as previously noted, there is a scale indetermination on
each component's spectrum when $\theta=\{ A, C_j(q), \sigma_d^2 \}$.  We
have found that this inherent indetermination must be explicitly fixed
in order for EM to converge (this is the rescaling step in the last
line of the pseudo-code).  Our strategy is, after each EM~step, to fix
the norm of each column of $A$ to unity and to adjust the
corresponding power spectra accordingly.  This is an arbitrary choice
which happens to work well in practice.

\subsection{Non linear optimization}\label{sec:nlin-optim}

When applied to our data, the EM algorithm shows fast convergence in a
first phase and then enters a second phase of slower convergence.
This is due to the fact that some parameters (\emph{e.g.} sub-dominant
power spectra in some spectral domains) have a very small effect on
the criterion.  In order to reach the true minimum of
$\phi(\theta)$, it appears necessary to complement EM with another
minimization technique.  The strategy is to use the straightforward EM
algorithm to quickly get close to the minimum of $\phi(\theta)$ and
then to complete the minimization using a dedicated minimization
algorithm.  This complementary algorithm can
use a simple design thanks to the good starting point provided by
EM.

The spectral mismatch criterion (\ref{eq:obj}) can, in theory, be
minimized by any optimization algorithm.  However, the same effect
which slows down EM in its final steps also makes the minimization of
the mismatch criterion (\ref{eq:obj}) difficult for \emph{any}
algorithm.  In particular, simple gradient algorithms are unacceptably
slow.  
Actually, we found that even conjugate gradient techniques cannot
overcome this problem and had to resort to a quasi-Newton method.  We
have used the classic BFGS (Broyden-Fletcher-Goldfarb-Shapiro)
algorithm~\citep{Luenberger73}.  This technique minimizes an objective
function by successive one-dimensional minimization (line searches).
At each step, the direction for the line search is the gradient
`rectified' by the inverse of Hessian matrix.  The BFGS technique is a
rule to update an estimate of the inverse Hessian matrix at low
computational cost.

\section{Testing and Performance} \label{sec:testing}

We now turn to illustrating the applications and performance of our
multi--detector multi--component spectral--matching method on a simple
set of synthetic observations: three-component noisy linear mixtures
featuring contributions from CMB anisotropies, dust emission, and SZ
thermal emission. Unbiasedness and statistical uncertainties are
investigated by a Monte--Carlo technique.

Five implementations of the method for different applications will be discussed:
\begin{enumerate}
  \item a multi--component spatial power spectrum estimation assuming
  the mixing matrix is known,
  \item a blind approach in which spatial power spectra, noise levels,
  and the emission laws of components are jointly estimated on the
  data,
  \item a semi--blind approach where CMB and SZ emission laws are
  assumed to be known, and the emission law of the dust component (in
  addition to spatial power spectra and noise levels for all
  components) is estimated from the data,
  \item an application for detector cross--calibration,
  \item a Wiener--filter component separation using parameters
  estimated via blind spectral matching.
\end{enumerate}
Finite beam effects are neglected for the present work, although they
are not a fundamental limitation for our method (see sec.
\ref{sec:discussion}). For definiteness, we also assume here that the
noise is white, although this assumption can be relaxed as well if
needed.

\subsection{Simulated data}

Synthetic observations in six frequency bands identical to those of
the Planck HFI are generated on $300 \times 300$ pixel maps
corresponding to a 12.5$\degree$ $\times$ 12.5$\degree$ field located
at high galactic latitude. For each mixture realisation, synthetic
components and noise are obtained as follows:
\begin{itemize}
  \item The CMB component is a COBE-normalised, randomly generated
  realization of CMB anisotropies obtained using the spatial power
  spectrum $C_\ell$ predicted by the CMBFAST software
  \citep{2000ApJS..129..431Z} with $H_0 = 65$ km/s/Mpc, $\Omega_m =
  0.3$, $\Omega_b = 0.045$, $\Lambda = 0.7$.
  \item The galactic dust emission template is obtained from the 100
  $\mu$m IRAS data in the sky region located around $\alpha =
  204\degree$ and $\delta = 11\degree$.  Bright stars are removed
  using a point source extracting algorithm.  Residual stripes are cut
  out by setting to zero the contaminated Fourier coefficients.  The
  Fourier modes suppressed in this way are randomly re-generated with
  a distribution obtained, for each mode, from the statistics of the
  other modes at the same scale in the IRAS map.  This method
  preserves the (assumed) statistical azimuthal symmetry and general
  shape of the spatial spectrum.
  \item The thermal Sunyaev-Zel'dovich template is drawn at random
  from a set of 1500 SZ maps generated for this purpose using the
  software described in \citep{delabrouille-melin-bartlett}.
  \item White noise at the level of the nominal per--channel Planck
  HFI values is added to the observations.
\end{itemize}

\begin{figure}
  \begin{center}     
    \includegraphics[width=\columnwidth]{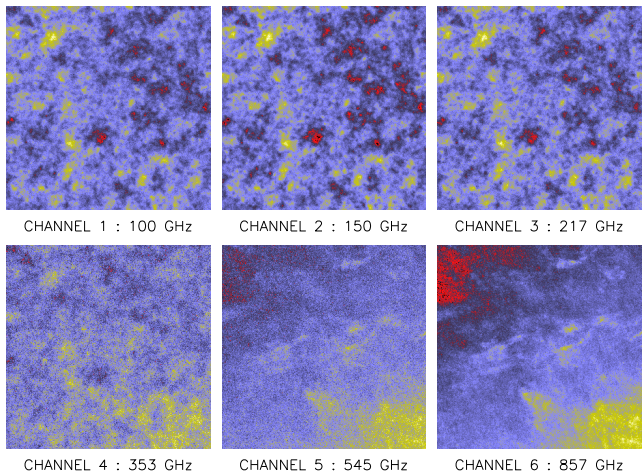}
  \end{center}
  \caption{Simulated observations for six frequency bands
  }
  \label{fig:allobs}
\end{figure}

Synthetic observations are displayed in fig.~\ref{fig:allobs}. The general 
common pattern which can be seen in the lowest frequency channels is 
simulated CMB anisotropies, whereas the pattern of emission of 
interstellar dust as observed with IRAS dominates our 857 and 545 GHz 
maps. The contribution of the SZ effect, very sub-dominant, is not 
obviously visible on these maps.

\begin{table}
  \begin{center}
    \tiny{
    \begin{tabular}{@{}lcccccc}
      $\nu$      & 100      & 143      & 217      & 353 & 545      & 857 \\ \hline
      CMB          & 0.889    & 0.926    & 0.896    & 0.275 & 0.0019 & 1.3$\times 10^{-7}$ \\  \hline
      dust         & 9$\times 10^{-5}$ & 6$\times 10^{-4}$ & 0.0082 & 0.215 & 0.687    & 0.938 \\ \hline
      SZ   & 0.0064   & 0.0032   & 2$\times 10^{-7}$ & 0.0044 & 0.00019 &  5.2$\times 10^{-8} $ \\ \hline
      noise        & 0.102   & 0.0727 & 0.108  & 0.536 & 0.320    & 0.0667 \\ \hline
    \end{tabular}
    }
  \end{center}
  \caption{Fraction of the power in each of the components} 
  \label{tab:fraction-all-channels}
\end{table}
Table~\ref{tab:fraction-all-channels} gives, for each channel, the
relative power of all components and of noise for a typical synthetic
mixture (here `relative' means: the sum of all powers is normalised to
unity). Typical input templates for the three components can be seen
in figure~\ref{fig:reconstructed-maps}, left column.

\subsection{Application 1: Spectral estimation}\label{application1}

The first application is the estimation of component spatial power
spectra. It is assumed that the mixing matrix is known, but that the
noise level for each map is not known precisely. The set of parameters
to be estimated from the data then is $\theta=\{C_j(q),\sigma_d^2\}$.

Component spectra are estimated on 32 ring-shaped domains for 5,000
different mixtures.  The first 30 domains are equally spaced rings
covering the lowest 60\% of the spatial frequencies ($0<
\ell/\ell_{\rm max}< 0.6$), and the remaining two cover respectively
$0.6< \ell/\ell_{\rm max}< 0.8$ and $0.8< \ell/\ell_{\rm max}< 1$.
This choice of spectral domains is adapted to the assumed azimuthal
symmetry of the spectra by the choice of ring-shaped domains, and has
a large number of rings in the region where the signal is strong and
where information from source spectra is relevant.

\begin{figure*}
        \begin{center}
        \includegraphics[width=2\columnwidth]{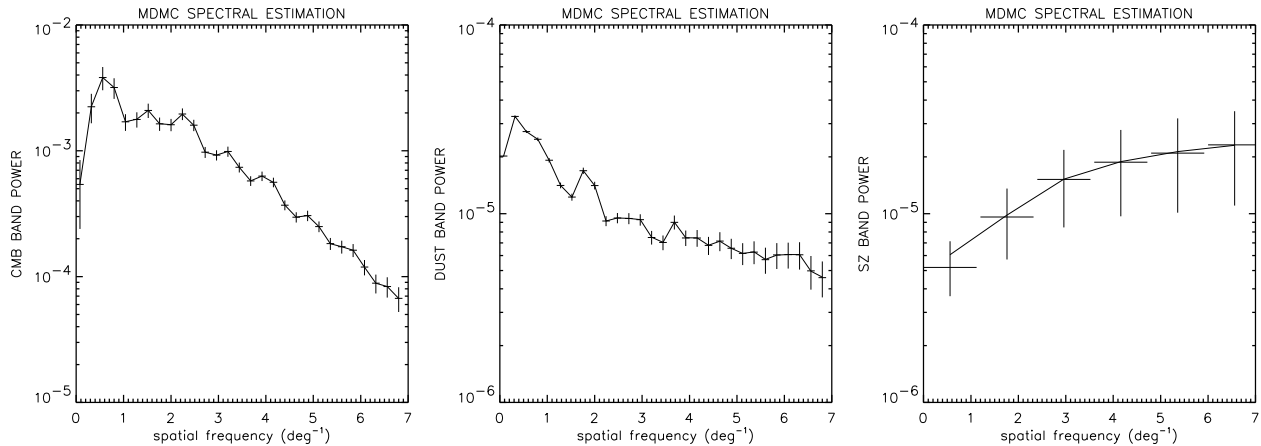}
        \end{center}
        \caption{This figure shows the recovered spatial power spectrum
        $\ell^2 C_\ell$ (crosses) compared to the exact band-averaged
        spectra (solid lines) for CMB (left), dust (middle), and
        Thermal SZ effect (right). These results correspond to a
        non-blind MDMC spectral estimation in which the mixing matrix
        $A$ is known.  Vertical bars show the $1\sigma$ errors, and
        horizontal bars the spatial frequency range of each bin.}
        \label{fig:spectral-est}
\end{figure*}

The result of the estimation of the spatial power spectrum of the
three components in the relevant frequency range is shown in
fig.~\ref{fig:spectral-est}. Errors on estimated spectra are obtained
from the dispersion over the 5,000 distinct simulated observations.
For the SZ effect, the spatial power spectrum is averaged into larger
bins after parameter estimation to reduce the scatter of the
measurements. The figure shows that, as expected, a low-variance
unbiased power spectrum is obtained for all components without
explicit separation of the observations into component maps. For the
CMB, the measurement is sample (cosmic) variance limited at small
spatial frequencies. Such an effect does not appear on the dust
spectrum estimate because we use only one dust map in the Monte-Carlo.

\subsection{Application 2: Blind parameter estimation}
\label{application2}

Let us now assume that the exact emission laws of all components are
unknown. Then the full parameter set, to be estimated from the data,
is $\theta = \{ C_j(q),\sigma_d^2,A \}$.  Again, we estimate
parameters on 5,000 different simulated data sets.  For each run, the
scale indetermination between mixing matrix columns and component
power spectra is fixed by renormalising to the true value of $A$ at a
single reference frequency (100 GHz for the CMB and thermal SZ effect,
and 857 GHz for the dust). Error bars ($\pm 1 \sigma$) for all
parameters are computed from the distribution of the estimates over
all simulated observations.

\begin{table*}
  \begin{minipage}{2\columnwidth}
    \begin{center}
      \begin{tabular}{@{}lcccccc}
        channel & 100 & 143  & 217  & 353 & 545 & 857\\
        \hline
        RMS est. $\times 10^{-6}$ & (29.1 $\pm$ 0.22)& (18.7 $\pm$ 0.13)& (12.85 $\pm$ 0.09)&  (11.92$\pm$ 0.07)& (8.98 $\pm$ 0.05)& (4.97 $\pm$ 0.06)\\
        \hline
        RMS true  $\times 10^{-6}$ & 29.11 & 18.70 &  12.86 & 11.93 & 8.980 & 4.970 \\
        \hline
      \end{tabular}
    \end{center}
  \end{minipage}
  \caption{
  Comparison of true and estimated noise levels (RMS). The errors
  are obtained from the dispersion of results obtained using 10,000
  different mixtures. }
  \label{tab:Noise-est}
\end{table*}
\begin{figure}
  \begin{center}
    \includegraphics[width=\columnwidth]{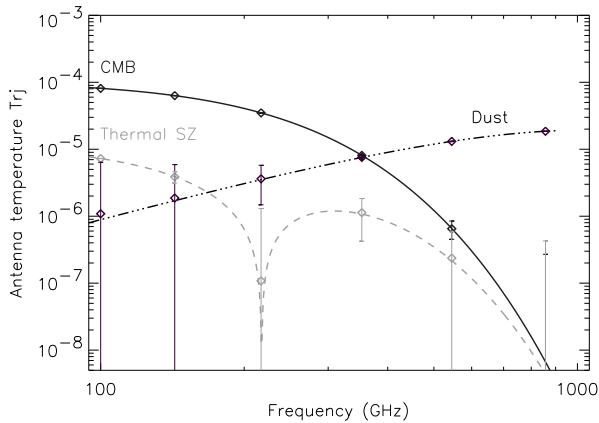}
  \end{center}
  \caption{
  The figure shows the recovered emission laws of the components
  (diamonds) compared to the exact emission laws used in the
  simulations (solid lines). The errors are computed from the
  dispersion of the recovered values for 10,000 different synthetic
  mixtures.  }
  \label{fig:res-a-out}
\end{figure}

Figure~\ref{fig:res-a-out} displays recovered emission spectra
(diamonds with 1$\sigma$ error bars) as compared to exact emission
spectra (solid lines).  Emission laws of all components are recovered
with no significant bias.  The CMB emission law is recovered very
accurately at all frequencies except 857 GHz.  The dust emission law
is recovered quite accurately at high frequencies, less accurately at
frequencies where it is very sub--dominant.  The SZ effect emission
shape, sub--dominant at all frequencies, is recovered with larger
relative error bars.  Because of the renormalisation, error bars for
CMB and SZ vanish at 100 GHz, and the dust emission law error bar
vanishes at 857 GHz.

\begin{figure*}
  \begin{center}
    \includegraphics[width=2\columnwidth]{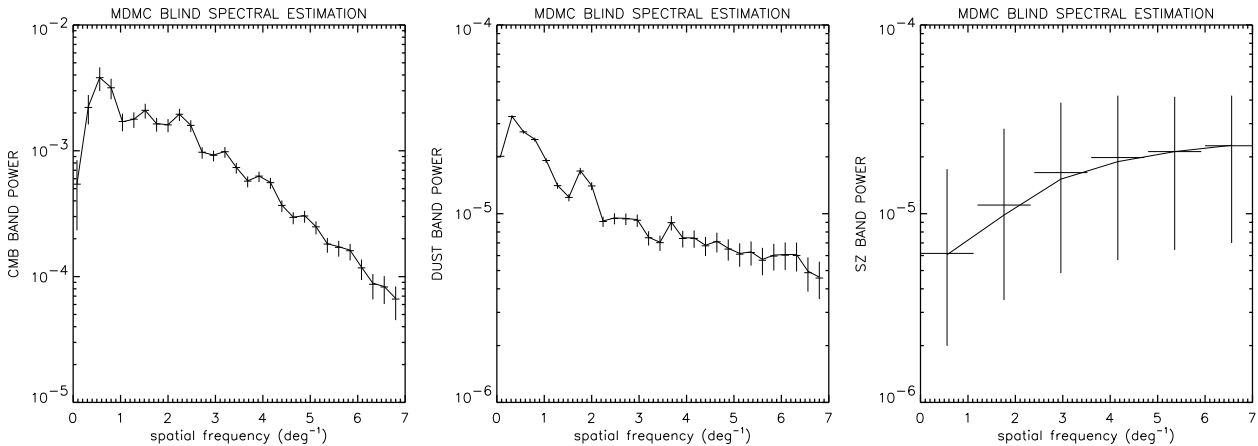}
  \end{center}
  \caption{
  The figure ---similar to fig.~\ref{fig:spectral-est} but for
  \emph{blind} spectral matching--- shows the recovered power spectra
  of the components compared to exact ones (solid lines). The errors
  are computed from the dispersion of the recovered values for 5,000
  different synthetic mixtures.  }
        \label{fig:blind-spectral-est}
\end{figure*}

Spatial power spectra, in turn, are also estimated.  As shown in
figure~\ref{fig:blind-spectral-est}, CMB and dust spatial power
spectra are recovered with good accuracy and no significant bias,
almost as well as for the non--blind spectral estimation. The SZ power
spectrum is also significantly constrained, although error bars are
significantly larger than in the non-blind spectral estimation.

Finally, table~\ref{tab:Noise-est} shows the estimates of the noise
RMS as compared to true levels.  Relative errors are below 2.5 \% for
all channels.

\subsection{Application 3: Semi-blind parameter estimation}
\label{application3}

In our particular case, the emission laws of the CMB and of the SZ are
known to almost perfect accuracy.  Assume, however, that measuring the
dust emission law is of particular interest.  How much do we gain by
forcing known emission laws to their true value, and estimating only
the unknown dust emission spectrum?

\begin{table*}
 \begin{minipage}{2\columnwidth}
   \begin{center}
     \begin{tabular}{@{}lcccccc}
       channel & 100 & 143  & 217  & 353 & 545 & 857\\
       \hline
       true dust em spectrum & 0.3071 & 0.5902 & 1.2177 & 2.6106 & 4.5371
       & 6.4288 \\
       \hline
       relative error, blind approach & 6.229 & 2.469 & 0.634 & 0.0662
       & 0.00790 &  no values \\
       \hline
       relative error, semi-blind approach & 2.623 & 1.056 & 0.285 & 0.0368
       & 0.00725 & no values \\
       \hline
     \end{tabular}
   \end{center}
 \end{minipage}
 \caption{Relative errors on dust emission law 
 estimation.  In the first case, all the elements of the mixing 
 matrix are estimated (blind approach).  In the second case, the 
 columns of the mixing matrix which corresponds to the CMB and the 
 thermal SZ components are fixed (semi-blind approach). Although the 
 semi-blind approach does not improve significantly the 
 determination of the dust spectrum at 545 GHz, the improvement is 
 very significant (factors of two to three) at other frequencies.}
 \label{tab:compare-dust-recov}
\end{table*}

We repeat the simulations described in~\ref{application2}, now fixing
two columns of the mixing matrix, and estimating the third one (in
addition to domain-averaged spatial spectra and noise levels). Table
\ref{tab:compare-dust-recov} compares quantitatively the relative
errors on the resulting dust emission law.  At low frequency (between
100 and 217~GHz), the accuracy of the estimation is improved by a
factor of 2 to 3.  At 353~GHz, the improvement is still noticeable,
but at 545~GHz, where the dust emission begins to dominate, the blind
and semi--blind approaches give similar errors.  The use of partial
prior information on the mixing matrix $A$ is thus useful here to
improve the estimation of the entries of $A$ which contribute little
relative power to the observations.

\begin{figure*}
  \begin{center}
    \includegraphics[width=2\columnwidth]{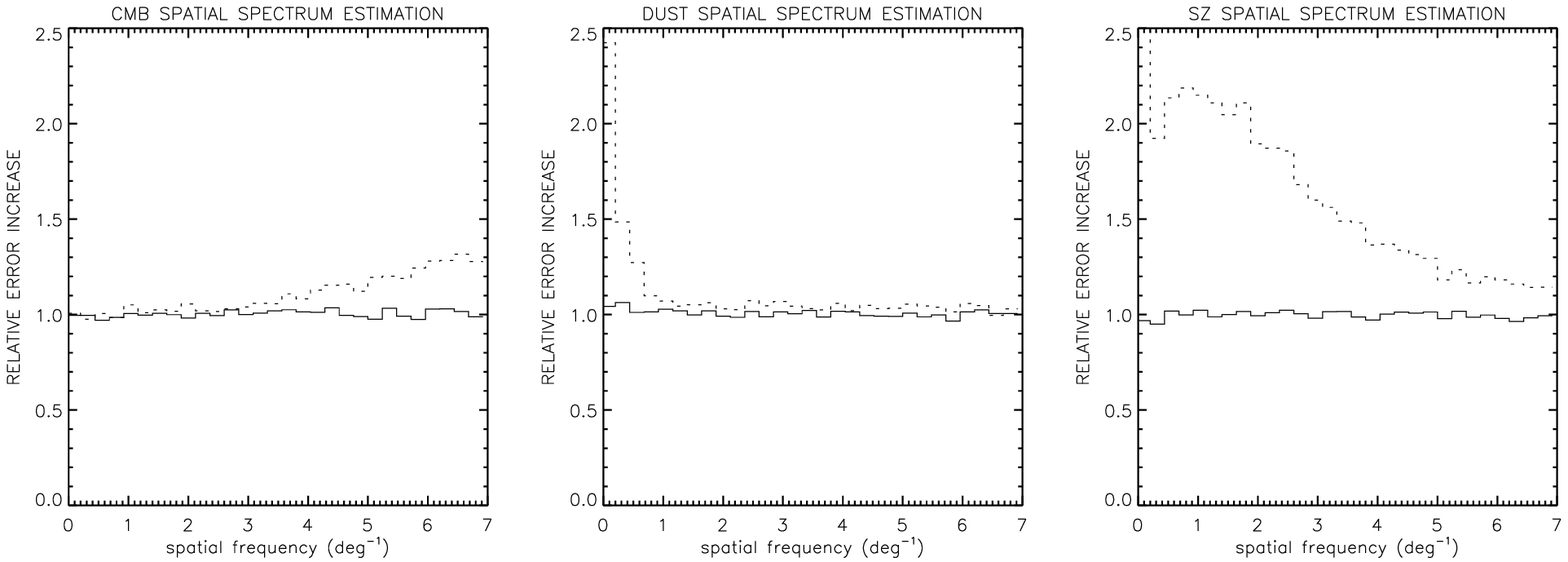}
  \end{center}
  \caption{
  The figure shows the comparison of the quality of the blind and
  semi-blind power spectrum estimation of the components.  The solid
  line displays the ratio between the size of the $1\sigma$ error in
  the semi--blind and in the non--blind spectral matching, showing
  that they are comparable to within simulation accuracy. In contrast,
  the dotted line shows the ratio between the size of the $1\sigma$
  error in the blind and in the non--blind spectral matching, showing
  that some accuracy is lost when all components of the mixing matrix
  are adjusted as additional parameters.}
  \label{fig:compare-spectral-est}
\end{figure*}
In addition to this substantial improvement in estimating the unknown
`dust column' of $A$, the semi--blind approach is more efficient for
estimating the SZ power spectrum than the full blind implementation.
Figure~\ref{fig:compare-spectral-est} shows the comparison of the
quality of spectral estimation in the blind and semi--blind approaches
relative to the non-blind. To the precision of our Monte-Carlo tests
(1--2\% level on error bars), the semi--blind result is as accurate
for this particular mixture as the non-blind estimate, significantly
better than the blind result. As the semi--blind and the non--blind
estimates give similar results, however, the actual enhancement in
precision depends on details of the mixture and parametrization.

This comparison, however, shows that it is in general useful to
exploit as much as possible reliable prior information. Our method is
flexible enough to do so.

\subsection{Application 4: Detector calibration}
\label{application4}

The mixing matrix $A$ depends not only on components (through emission
spectra), but also on detectors (through frequency bands and optical
efficiency). Mixing matrix coefficients $A_{dj}$, expressed in readout
(rather than physical) units can be approximated by the product of a
detector-dependent calibration coefficient $\alpha_d$ and an emission
law~$\epsilon_j(\nu)$:
\begin{equation}
  A_{dj} \simeq \alpha_d \epsilon_j(\nu_d)
\end{equation}
where $\nu_d$ is the central observing frequency of detector $d$.
Used on a data set from detectors observing in the same frequency
band, the estimation of $A$ for any astrophysical component gives
relative calibration coefficients between detectors. If in addition
the emission law of at least one of the components is known (e.g. CMB
anisotropies), the estimation of the mixing matrix provides a relative
calibration across frequency bands. Finally, if among the components
there is one with known emission spectrum and known amplitude (or
known spatial power spectrum), absolute calibration can be obtained in
the same way. For instance, it is not excluded that in the not-so-far
future, a high resolution experiment dedicated to a wide-field point
source survey in the millimeter range can be calibrated on CMB
anisotropies(!).

\subsection{Application 5: Component separation}
\label{application5}

The separation of astrophysical components by some kind of inversion
of the linear system of equation~\ref{eq:mixture} has been the object
of extensive previous work. Popular linear methods are listed in
appendix~\ref{sec:sepcomp}. In a Gaussian model, the best inversion is
obtained by the Wiener filter. This filter, however, requires the
prior knowledge of the mixing matrix $A$, component spatial power
spectra, and noise power spectra. As discussed by
\citep{cmb:eusipco02}, our spectral--matching method yields all the
parameters needed to implement a Wiener--based component separation
\emph{on maps}.

We compare the quality of component reconstruction using either the
estimated parameter set $\theta = \{ C_j(q),\sigma_d^2,A \}$ or `true'
best-knowledge values.

\begin{figure*}
  \begin{center}
    \includegraphics[width=2\columnwidth]{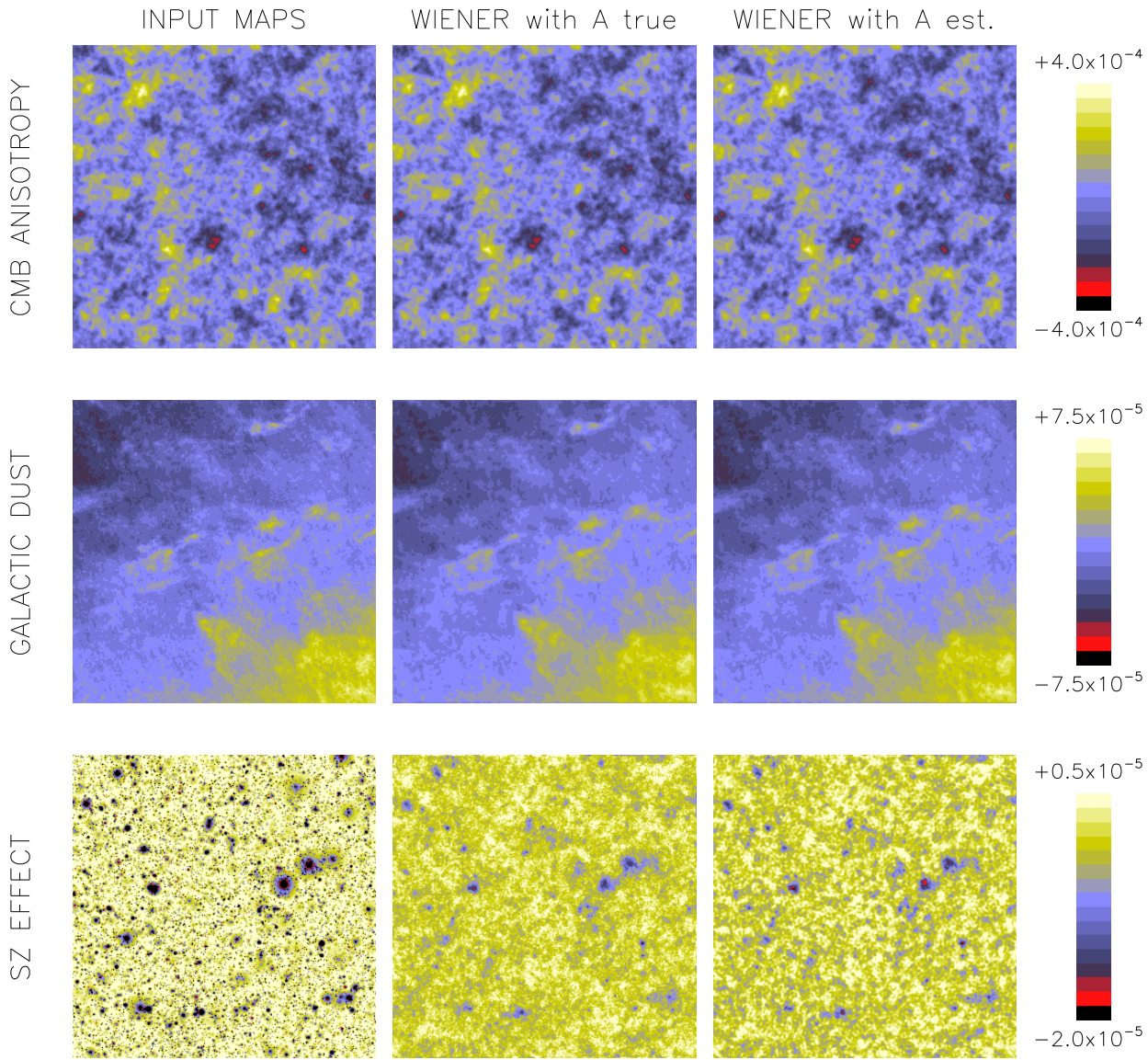}
  \end{center}
  \caption{
  Left: true templates used as inputs. Middle: templates recovered
  using the `true' Wiener filter. Right: templates recovered using the
  blind separation.  }
  \label{fig:reconstructed-maps}
\end{figure*}

\mypar{Reconstructed maps.}
Figure~\ref{fig:reconstructed-maps} illustrates the quality of map
reconstruction by Wiener inversion. The first column displays the
input components, the second column shows components recovered with
the exact Wiener filter (computed from the true mixing matrix,
ensemble averages of the noise, ensemble averages of CMB and SZ power
spectra, and a $k^{-3}$ fit of the spatial power spectrum of the dust
template). The third column displays the components recovered by
Wiener inversion using estimated parameters.  In both cases, CMB and
dust emissions are recovered satisfactorily, but the SZ effect
---strongly peaked and hence poorly suited to processing in Fourier
space--- remains noisy.  Visually, both methods perform about as well.

\mypar{Contamination levels.} 
The quality of the separation can be assessed by a measure of
contamination levels, i.e. how much of the other components gets into
a component's map after separation.

The Wiener matrix, $W=[A^tR_N^{-1}A + R_{S}^{-1}]^{-1} A^tR_N^{-1}$, 
obtained with exact values of $A$, $R_N$ and $R_{S}$, differs slightly 
from its estimate $\widehat W$, computed with estimates ${\widehat 
A}$, ${\widehat {R}_N}$ and ${\widehat {R}_{S}(q)}$.  
Not only ${\widehat {R}_{S}(q)}$ differs from $R_{S}$
because it is an estimate, but also because it is a flat band-power 
approximation.

\begin{figure*}
  \begin{center}
    \includegraphics[width=2\columnwidth]{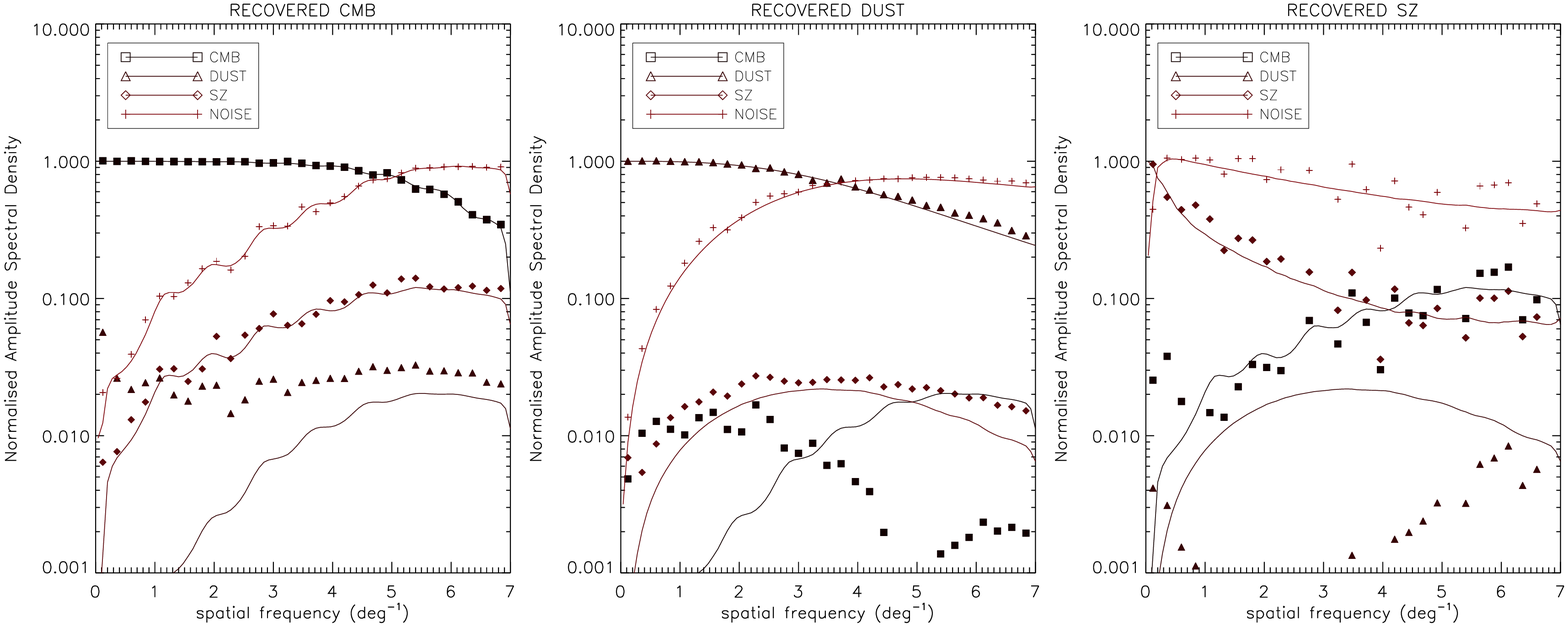}
  \end{center}
  \caption{
  Contributions to the output map as a function of spatial frequency,
  relative to the true level of that component map. The left panel is
  for CMB, the middle panel for dust, and the right panel for the SZ
  effect. Results obtained with exact values of the mixing matrix, the
  spectra and noise levels are plotted as plain lines, and results
  obtained with the Wiener implementation using estimated parameters
  as diamonds.}
  \label{fig:contamin-after-sep}
\end{figure*}

At each frequency, off-diagonal terms of $\widehat W A$ correspond to
leakage of other components into one component's estimate at spatial
frequency $k$.  Each panel of figure~\ref{fig:contamin-after-sep}
refers to one component (CMB, dust and SZ), and shows the relative
contribution of all components and of noise to the recovered map.
Levels are relative to the true map, so that the contribution of a
component to its own recovered map illustrates the spatial filtering
induced by the Wiener inversion.  The figure illustrates that the
inversion done with blindly estimated parameters performs almost as
well as the separation using exact values of the spectra and mixing
matrix. Differences are typically much smaller than noise
contamination, which is comparable with the blind and the non-blind
approaches.

\section{Discussion}
\label{sec:discussion}

\subsection{Related work on component separation}

Explicit component separation has been investigated first in CMB
applications by \citet{1996MNRAS.281.1297T},
\citet{1999NewA....4..443B}, and \citet{1998MNRAS.300....1H}. In these
applications, all the parameters of the model (mixing matrix, noise
levels, statistics of the components, including the spatial power
spectra) are assumed to be known.
  
Recent research has addressed the case of an imperfectly known mixing
matrix.  It is then necessary to estimate it (or at least some of its
components) directly from the data.  For instance, Tegmark \textit{et
al.} assume power law emission spectra for all components except CMB
and SZ, and fit spectral indices to the observations
\citep{2000ApJ...530..133T}.
  
More recently, it has been proposed to resort to `blind source
separation' or `independent component analysis' (ICA) methods.  The
work of \citet{2000MNRAS.318..769B}, further extended by
\citet{2002MNRAS.334...53M} implements a blind source separation
method exploiting the non--Gaussianity of the sources for their
separation.  This infomax method, unfortunately, is not designed for
noisy mixtures and can not deal with a frequency--dependent beam.
  
The idea to use spectral diversity and an EM algorithm for the blind
separation of components in CMB observations was proposed first by
\cite{snoussi}. This approach exploits the spectral diversity of
components as in our MDMC spectral matching, but assumes the prior
knowledge of the spatial power spectra of the components.  Our
approach extends further on this idea, with a lot more flexibility,
and the new point of view that spatial power spectra are actually the
main unknown parameters of interest for CMB observations.
  
Other reports of blind component separation in astronomical data
include~\cite{NuzBijAASS} and~\cite{valpola:ica2001}.

\subsection{Comments on the spectral matching approach}

\mypar{Robustness}

Our approach assumes that the data are collected in the form of a
linear mixture of a known number of components that are independent,
have different spatial power spectra, and different laws of emission
as a function of frequency.  These assumptions are valid in the
three-component mixtures used in our simulations. Applying this method
to real data obtained with the Archeops
experiment~\citep{archeops-c-ell} gave us the opportunity to test that
the method is quite robust, with satisfactory performance even when
the noise is not white nor stationary, and when some residual
systematic effects remain in the data. Of course, the exact impact of
large departures from the model remains to be tested on a case by case
basis.

\mypar{Detector--dependent beams}

It is quite usual in CMB observations that, because of the diffraction
limit, the resolution of the available maps depend a lot on frequency.
For Planck, the resolution ranges from about 30 arc-minutes at 30 GHz
to 5 arc-minutes at 350 GHz and higher. It is mandatory that a method
combining all observations can benefit from the full resolution of the
highest frequency channels. MDMC spectral matching, being implemented
in Fourier (or spherical harmonics) space, permits to take beam
effects into account straightforwardly by including in the model the
effect of a transfer function.

\mypar{Identifying components}

In practice, MDMC spectral matching runs with a fixed number of
components.  This number might not be well known (or even not very
well defined), and must be guessed (or assumed).  For CMB
applications, an educated guess can be made (as usual for all
component separation methods).

A practical way to handle this issue consists in applying the method
several times with a increasing number of expected components.
Comparing successive results permits to identify `stable' components,
which remain essentially unchanged when more components are sought.
Too few components result in unsatisfactory identification and poor
adjustment of the model to the empirical spectrum. Too many components
results in the separation of artificial components, either very weak,
or single detector noise maps.

With this strategy, the method can be seen as a component
\emph{discovery} tool, which can be useful in particular to uncover
and separate out instrumental effects behaving as additional
components.

Connected to the issue of component identification is the uniqueness
(or identifiability) problem.  As discussed above, MDMC spectral
matching uses spectral diversity as the `engine' of blind separation:
components with proportional spatial power spectra (or nearly so) are
not (or poorly) separated.  In the current test, the three components
are different enough that no such problem arises.  In richer mixtures,
containing contributions from several galactic components, it is quite
possible that spectral diversity does not hold. If, for instance,
several galactic components have a spatial power spectrum proportional
to $1/k^3$, the method would satisfactorily estimate parameters
relevant to the CMB and the SZ effect, but is unable to unmix galactic
contributions. A way out is to use a semi--blind approach in which
some entries of the mixing matrix are forced to zero when the
contribution of a particular component at a particular frequency is
known to be negligible. This is the object of forthcoming research.

\subsection{Comments on the Wiener inversion}

After adjusting the parameters of the model to the data, the recovered
mixing matrix, spectra, and noise levels can be used for component
separation by Wiener inversion.

Quite interestingly, the Wiener filter can be implemented for
identified components even if some sub-mixtures are not identified
(for instance by lack of spectral diversity). It can be shown
straightforwardly that the Wiener form:
\begin{equation}
  W=[A\adj R_N^{-1}A + R_S^{-1}]^{-1} A\adj R_N^{-1}
\end{equation}
can be rewritten equivalently as: 
\begin{equation}
  W=R_SA\adj [AR_SA\adj  + R_N]^{-1}
\end{equation}
or 
\begin{equation}
  W=R_SA\adj R_Y^{-1}
\end{equation}
Thus, the Wiener inversion for component $j$ requires only an estimate
of $R_Y$ (readily available as $\widehat R_Y$), of the spatial power
spectrum of component $j$, and of the column of the mixing matrix $A$
corresponding to component $j$. Therefore, \emph{it is not necessary
to identify all components, nor to know all spatial power spectra, nor
to know noise levels, to separate the CMB from the other components}.
We just need to know the CMB emission law (which we do) and its
spatial power spectrum (which can be estimated blindly with our
method).

As a final note, we stress that the Wiener method has the property of
filtering the data spatially -- an unpleasant fact when power spectra
are estimated on separated maps. In contrast, MDMC spectral matching
adjusts domain--averaged spatial power spectra on the data prior to
component map separation (bypassing the need for power--spectrum
estimation on output maps).

\subsection{Comments on spectral estimation}

In the above discussion, we have assumed for simplicity that the noise
is spatially white for all detectors. This assumption, however, can be
relaxed if needed, without (in general) loosing identifiability.

If the noise is uncorrelated between detectors, noise terms appear
only on the diagonal of the multivariate power spectrum of the
observations $R_Y$. Off--diagonal terms contain only contributions
from the off--diagonal terms of $AR_SA\adj$. If noise power spectra
are completely free, off-diagonal terms of $\widehat R_Y$ constrain
$AR_SA\adj$, and diagonal terms serve to measure $R_N$.

For instance, if the mixing matrix $A$ is known, it is possible to
adjust simultaneously the spatial power spectrum of the components and
that of the noise on the data, as long as enough observations are
available which is generically the case.

If data from several experiments are analyzed jointly, however, no
correlated noise of instrumental origin is expected between data from
detectors belonging to different experiments. This provides strong
consistency checks, which ultimately provides an additional handle on
the assessment of errors in the final results.

With a MDMC approach in Fourier (or spherical harmonic) space, data at
different frequencies and with different beam sizes can be analyzed
jointly. This joint analysis can be done straightforwardly by stacking
all observations from different instruments in the same vector of
observations $y$, as long as they cover the same area of the sky.
This is bound to become of major importance for the future scientific
exploitation of multi-scale and multi-frequency data.

\subsection{Using single detector maps}

For a well calibrated instrument, the linear mixture model can be
written in physical units, and the mixing matrix $A$ depends only on
the emission laws of components. Traditionally then, component
separation is implemented on a set of maps per frequency channel (data
from all detectors in each single frequency channel are combined into
a single map). This approach should be preferred if good maps cannot
be obtained independently for each detector (for sampling reasons, or
because of striping\dots), and if all detector data at the same
frequency can be combined (with some optimality) into one single map.

An alternate solution, when calibration coefficients and noise
properties for individual detectors (levels, correlations between the
noise of different detectors) are not known precisely, is to estimate
parameters directly using single detector maps in readout units (e.g.
microvolts), which can be done naturally with our spectral--matching
method.

\subsection{Comment on domain averaging}

We have considered band-averaged spectra as in
definition~(\ref{eq:Cjq}).
In CMB studies, one may be more interested in quantities like
$\ell(\ell+1) C_j(\vec\ell)$ which are expected to vary more slowly
than $C_(\vec\ell)$ itself.
In this case, it may be more appropriate to perform bin averages as
\begin{equation}
  \tilde R_Y(q) 
  = \left( 
  \sum_{\vec\ell\in\dom_q} \ell(\ell+1) \right)\inv
  \sum_{\vec\ell\in\dom_q} \ell(\ell+1) \ Y(\vec \ell)  Y(\vec \ell)\adj
  .
\end{equation} 
Spectral matching on such statistics would then yield estimates of
\begin{equation}
  \tilde C_j(q) 
  =
  \left( \sum_{\vec\ell\in\dom_q} \ell(\ell+1) \right)\inv
  \sum_{\vec\ell\in\dom_q} \ell(\ell+1)  C_j(\vec\ell)
  .
\end{equation} 
This weighted band-averaging can be used in our MDMC
spectral--matching method as well.

\section{Conclusion}\label{sec:conclusion}

This paper describes a spectral matching method for blind source
identification in noisy mixtures.  The method adjusts a simple model
of the data to the observations.  We estimate a physically relevant
set of parameters (fundamental parameters of the model: the mixing
matrix, domain-averaged spatial power spectra of the sources and of
noise) by maximum likelihood.  Only unknown parameters are estimated,
as the method lends itself easily to the modifications necessary to
exploit partial prior information.  Thanks to a Gaussian stationary
model, the likelihood depends only on a reduced set of statistics
(average spectral density matrices of the observations). An efficient,
dedicated algorithm can adjust the parameters in just a few minutes on
a modest workstation.

Our method is of particular relevance for CMB data analysis in a
multi-detector, multi-channel mission as Planck.

First, the method permits the blind separation of underlying
components, hence, of emissions coming from different astrophysical
sources.  Obtaining clean maps of emissions due to distinct
astrophysical processes is crucial to understanding their properties.

Second, the blind method permits to estimate the number of components
(by repeating the adjustment with a varying number of sources).  This
will be of utmost importance for analysing data from sensitive
missions as Planck, in particular for the identification and
characterisation of sub-dominant processes of foreground emission
(e.g. free-free emission, non-thermal dust emission), or to track down
systematic effects in the data.

Third, the blind method can estimate the entries of the mixing matrix.
This permits, if needed, to constrain the emission law
(electromagnetic spectrum) of the different components contributing to
the mixture, which is essential for understanding their physical
properties and possibly the emission processes.

Fourth, if strong sources, for which the mixing matrix is well
recovered, contribute to the mixture, the method can provide a useful
tool for the inter-calibration (or the absolute calibration) of the
different detectors or of the different channels.

Fifth, as our method is essentially a spectral matching method, which
adjusts the spectra of a number of components to the observational
data, it provides a direct measurement of the spatial power spectrum
of the components in the mixture, of particular relevance for the CMB.

As a final word, let us emphasize that the method can be applied to
sets of data coming from different experiments. As the MDMC spectral
matching approach, implemented in Fourier space, permits
straightforwardly to account for beam effects, it permits also to
analyze jointly and blindly multi--experiment, multi--channel,
multi--detector, multi--resolution data as long as they cover the same
area of the sky.  The method may become an essential tool for mapping
and analyzing sources of emission observed with present and upcoming
sub--millimeter experiments.

\section{Acknowledgements}

We acknowledge useful discussions with Mark Ashdown, Mike Hobson, Juan
Macias, Ali Mohammad-Djafari, Hichem Snoussi, and the Archeops
collaboration. This work was made possible by a grant from the French
ministry of research to stimulate inter-disciplinary work for applying
state-of-the-art signal and image processing techniques to CMB data
analysis.

\appendix

\section{Linear component separation}\label{sec:sepcomp}

The separation of astrophysical components relies on the key
assumption that the total sky emission at frequency $\nu$ is a linear
superposition of a number of components as in
equation~\ref{eq:mixture}.  In principle then, the observation of the
sky emission at several frequencies $(\nu_{1}, \nu_{2}, \ldots)$
permits to recover estimates $\widehat s_{j}(\theta, \phi)$ of the
component templates $s_{j}(\theta, \phi)$ by inverting
equation~\ref{eq:mixture}.  There are several methods for a linear
inversion of the system when the mixing matrix $A$ is known:

\begin{enumerate}
    
  \item If there are as many noiseless observations as there are
  astrophysical components contributing to the total emission, by
  simple inversion of the square matrix $A$, so that the recovered
  components, $\widehat S$, are given by $\widehat S = A^{-1} Y$;
    
  \item If there are more observations than astrophysical components,
  the system can be inverted using the pseudo inverse, $\widehat S =
  [A\adj A]^{-1}A\adj Y$;
    
  \item For optimal signal to noise ratio under Gaussian statistics,
  without other prior assumption on the astrophysical components, one
  can use a generalized least square solution, $\widehat S = [A\adj
  R_N^{-1}A]^{-1} A\adj R_N^{-1}Y$, where $R_N$ is the noise
  correlation matrix;
    
  \item The choice $\widehat S = [A\adj R_N^{-1}A + R_S^{-1}]^{-1}
  A\adj R_N^{-1}Y = WY$ is the Wiener solution.  It is the linear
  solution which minimises the variance of the error, but requires the
  knowledge of both the noise autocorrelation, $R_N$, and of the
  component autocorrelation, $R_S$.  As $[WA]_{ii} \leq 1$, this
  solution modifies the spatial spectra of the components since
  different weights are given to different spatial frequencies of a
  component map.
    
  \item The renormalised Wiener solution, $\widehat S = \Lambda WY$,
  where $\Lambda = [\diag (WA)]^{-1}$, is the Wiener solution under
  the constraint $[WA]_{ii} = 1$.  This solution renormalises the
  Wiener solution at each spatial frequency, so that no spatial
  filtering is applied to the data.
    
\end{enumerate}

In the above list, solution 1 is the special case of 2 when $A$ is
square and regular, 2 is the special case of 3 when the noise is white
$(R_N \propto {\rm Id})$, 3 the special case of 4 when the signal is
much stronger than the noise, and 5 a constrained version of 4 that
does not modify the relative importance of different spatial
frequencies in a component map after inversion.  Depending on the
method chosen, one or more of $A$, $R_N$ and $R_S$ (which can be
considered as parameters of the model) is needed to implement the
inversion.

Realizing the fact that optimal component separation requires the
prior knowledge of a set of parameters of the model is one of the
driving ideas of our MDMC spectral--matching approach: we implement
the \emph{joint estimation} of all such parameters that are not
necessarily known \emph{a priori}.

\section{Spectral matching and likelihood}\label{sec:whittle}

This section shows that minimizing the spectral matching 
criterion~(\ref{eq:obj}) is equivalent to maximizing the likelihood of 
a simple model.

\mypar{Gaussian likelihood and covariance matching}

We first show how criterion~(\ref{eq:obj}) is related to a Gaussian 
likelihood.  If $y$ is a real $n\times 1$ zero mean Gaussian random 
vector with covariance matrix $R$, then
\begin{equation}
  -2 \log p(y) = y\adj R\inv y + \log\det (2\pi R)
  .
\end{equation}
If $Y=[ y_1, \ldots, y_T ]$ is an $n\times T$ matrix made of $T$ such 
vectors, independent from each other, with $\mathrm{Cov}(y_t)=R_t$, 
then
\begin{equation}\label{eq:probY}
  -2 \log p(Y) = \sum_{t=1}^T y_t\adj R_t\inv y_t + \log\det (2\pi R_t)
\end{equation}
Assume further that the index set $[1,\ldots,T]$ can be decomposed in
$Q$ subsets $I_1,\ldots,I_Q$ such that $R_t$ is constant with value
$R(q)$ over the $q$th subset, that is, $R_t = R(q)$ if $t\in I_q$.
Then, eq.~(\ref{eq:probY}) can be rewritten, using $ y\adj R\inv y =
\trace\left( R\inv y y\adj \right)$ as
\begin{displaymath}
  \nonumber
  -2 \log p(Y) =
  \sum_{q=1}^Q n_q \left[\trace \widehat{R}(q) 
  \left(R(q)\inv \right) + \log\det (R(q)) \right] + \mathrm{cst}
\end{displaymath}
where $ \widehat{R}(q) = \frac 1 {n_q} \sum_{t\in I_q} y_t y_t\adj$ 
and $n_q$ is the number of indices in $I_q$.  This last expression 
also reads
\begin{equation}\label{eq:mlaskf}
  -2 \log p(Y) = \sum_{q=1}^Q n_q D\left(\widehat R_y (q), R_y(q) 
  \right) + \mathrm{cst}
\end{equation}
where the constant term is a function of the data $Y$ via 
$\widehat R_y(q)$ but not of any $R(q)$.  This form makes it clear 
that the mismatch~(\ref{eq:obj}) corresponds to the log-likelihood of 
a sequence of zero mean Gaussian vectors which are modeled as having 
block-wise identical covariance matrices.

\mypar{Whittle approximation}

The statistical distribution of the Fourier coefficients of a 
stationary time series is a well researched topic.
If $T$ samples $y(1),\ldots,y(T)$ of an $n$-variate discrete time 
series are available, the Fourier transform is:
\begin{equation}
  \widetilde y(f ) 
  =
  \frac 1 {\sqrt T} \sum_{t=0}^{T-1} y(t) \exp -2i \pi f  t
  .
\end{equation}
For a stationary time series with spectral covariance matrix
$R(f )$, simple asymptotic (for large $T$) results are available.
In particular, the Whittle approximation consists in approximating the
distribution of the Fourier transform $\widetilde y(f )$ at DFT points
$f = q/T$ as follows:
\begin{itemize}
  \item The real part and the imaginary part of $\widetilde y(f )$ are
  Gaussian, uncorrelated, with the same covariance matrix and 
  $E \widetilde y(f )\widetilde y(f )\adj = R(f )$.
  \item For $0<p \neq p'< T/2$ (assuming $T$ even and for $p,p'$
  integers), $\widetilde y(p/T)$ is uncorrelated with $\widetilde y(p'/T)$.
\end{itemize}
This is a standard approximation: it has been used for the blind
separation of noise free mixtures of components by
\cite{Pham:IEEESP:97} and in the context of astronomical component
separation by~e.g.\cite{1999NewA....4..443B,1996MNRAS.281.1297T}.

Expression~(\ref{eq:mlaskf}) thus shows\footnote{
Actually some care is required to deal with the fact that the Fourier
coefficients are complex-valued and that $\widetilde y(-f ) = \widetilde
y(f )^\star$.  This introduces some minor complications in the
computations but does not affect the final result.  }
that the minimization of~(\ref{eq:obj}) is equivalent to maximizing
(the Whittle approximation to) the likelihood provided we model the
spectra of the sources as being constant over spectral domains.

\section{An EM algorithm in the spectral domain}\label{sec:domain-em}

The Expectation-Maximization (EM) algorithm~\citep{Dempster77} is a
popular technique for computing maximum likelihood estimates.  This
section first briefly reviews the general mechanism of EM and then
shows its specific form when applied to our model.

\mypar{The EM algorithm.}
Consider a probability model $p(y,s|\theta)$ for a pair $(y,s)$ of
random variables with $\theta$ a parameter set.
If the variable $s$ is not observed, the log-likelihood of the
observed $y$ is
\begin{equation}
  \label{eq:defll}
  l(\theta) 
  =
  \log  p(y|\theta)
  = 
  \log \int p(y,s|\theta) ds
\end{equation}
For some statistical models, the maximization of the log-likelihood 
$l(\theta)$ can be made easier by considering the EM functional:
\begin{equation}
  \label{eq:defem}
  l(\theta,\theta') 
  =
  \int \log( p(y,s|\theta))\ p(s|y,\theta') ds
  .
\end{equation}
The EM algorithm is an iterative method which computes a sequence of
estimates according to:
\begin{equation}
  \label{eq:emstep}
  \theta^{(n)}
  \rightarrow
  \theta^{(n+1)}
  =
  \arg\max_\theta l\left(\theta, \theta^{(n)} \right)
\end{equation}
It can be shown that
\begin{equation}
  \label{eq:propema}
  l(\theta'',\theta') > l(\theta',\theta')
  \Rightarrow
  l(\theta'') > l(\theta')
\end{equation}
meaning that every step of the algorithm can only increase the
likelihood.
Actually, a stationary point of the algorithm also is a stationary
point of the likelihood since
\begin{equation}
  \label{eq:propemb}
  \frac{\partial l(\theta) }{\partial \theta }
  =
  \left.
    \frac{\partial l(\theta,\theta') }{\partial \theta } 
  \right|_{\theta'=\theta}
\end{equation}
The EM algorithm is an interesting technique for maximizing the
likelihood if
i) the computation of the conditional expectation in
definition~(\ref{eq:defem}) (E step) is and
ii) the maximization~(\ref{eq:emstep}) of the functional (M step)
are computationally tractable.

Both the E~step and the M~step turn out to be straightforward because
one elementary EM step amounts to solving:
\begin{equation}
  \label{eq:eqemeq}
  0 
  =
  \int \frac{\partial \log( p(y,s|\theta^{(n+1)})) }{\partial\theta }  \ p(s|y,\theta^{(n)}) \, ds
  .
\end{equation}
In our model, the partial derivative in~(\ref{eq:eqemeq}) turns out to
be a simple function of $y$ and $s$, allowing the conditional
expectation to be easily computed and eq.~(\ref{eq:eqemeq}) to be
easily solved.
This is sketched in the following.

\mypar{A single Gaussian vector.}
In order to introduce the necessary notations, we start by considering
a simple case where $y=As+n$ where $s$ and $n$ are independent
Gaussian vectors with zero-mean and covariance matrices equal to $R_s$
and $R_n$ respectively.  Then the parameter set is
$\theta=(A,R_s,R_n)$ and one has
\begin{eqnarray}
  \nonumber
  -2\log p(y|s,\theta) & = & (y-As)\adj R_n\inv(y-As) + \log|R_n| + \mathrm{cst} \\
  \nonumber
  -2\log p(s|\theta) & = & s\adj R_s\inv s + \log|R_s| + \mathrm{cst}
\end{eqnarray}
Using $p(y,s)=p(y|s)p(s)$, the log derivatives of the joint density
with respect to the components of $\theta$ are:
\begin{eqnarray}
  \label{eq:dlda} \frac{\partial\log p(y,s|\theta) }{\partial A }       &=& R_n\inv \left[ (y-As)s\adj \right] \\
  \label{eq:dldn} \frac{\partial\log p(y,s|\theta) }{\partial R_n\inv } &=& -\frac 1 2 \left[ (y-As)(y-As)\adj -R_n \right]    \\
  \label{eq:dlds} \frac{\partial\log p(y,s|\theta) }{\partial R_s\inv } &=& -\frac 1 2 \left[ ss\adj - R_s \right]
\end{eqnarray}
Thus, in this simple model, computing the conditional expectations as
in eq.~(\ref{eq:eqemeq}) would boil down to evaluating the conditional
expectations of the random variables $ss\adj$, $sy\adj$, $ys\adj$ and
$yy\adj$.
This is a routine matter in a Gaussian model $y=As+n$ for which one
finds:
\begin{eqnarray}
  \label{eq:esscy} E(ss\adj|y, \theta) &=& W(\theta)yy\adj W(\theta)\adj + C(\theta) \\
  \label{eq:esycy} E(sy\adj|y, \theta) &=& W(\theta)yy\adj \\
  \label{eq:eyscy} E(ys\adj|y, \theta) &=& yy\adj W(\theta)\adj \\
  \label{eq:eyycy} E(yy\adj|y, \theta) &=& yy\adj 
\end{eqnarray}
with the following definitions for matrices $C(\theta)$ and $W(\theta)$:
\begin{eqnarray}
  \label{eq:defC} C(\theta) &=& (A\adj R_n\inv A + R_s\inv )\inv \\
  \label{eq:defW} W(\theta) &=& (A\adj R_n\inv A + R_s\inv )\inv  A\adj R_n\inv
\end{eqnarray}
Note that $C(\theta)=\mathrm{Cov}(s|y, \theta)$ and that $W(\theta)$
is the Wiener filter, that is $E(s|y,\theta)=W(\theta)y$.

\mypar{The EM algorithm in the Whittle approximation}
In our model, according to the Whittle approximation,
the DFT points $y(k)$ are independent so that
the EM functional~(\ref{eq:defem}) for the whole data set simply is a
sum over DFT frequencies of elementary functionals.
Thus an EM step $\theta' \rightarrow \theta$ consists in solving
\begin{equation}
  \label{eq:emstepW}
    0 
    =
    \sum_k E
    \left\{ 
      \frac {\partial} {\partial\theta}
      \log  p 
      \left(
        y(k),s(k)|\theta
      \right) 
      |\ y(k), \theta'
    \right\}
    .
\end{equation}
To proceed further, eq.~(\ref{eq:emstepW}) is specialized to the case
of interest by using two ingredients.  
First, we use the relation $y(k)=As(k)+n(k)$ and the Gaussianity of
each pair $(y(k),s(k))$; this is expressed via
eqs.~(\ref{eq:dlda}-\ref{eq:dlds}).
Second, we use the approximation that the power spectra are constant
over each spectral domain.
Combining these properties, the cancellation~(\ref{eq:emstepW}) of the
gradient with respect to $A$, $R_n$ and each $R_s(q)$ yields
\begin{eqnarray}
  \label{eq:demda} 0&=&\widetilde R_{ys}(\theta') - A(\theta) \widetilde R_{ss}(\theta') \\
  \label{eq:demdn} 0&=&\widetilde R_{yy}(\theta') - A(\theta) \widetilde R_{sy}(\theta') - \widetilde R_{ys}(\theta')A(\theta) \adj \nonumber \\
                    & &  + A(\theta) \widetilde R_{ss}(\theta')A(\theta) \adj - R_n \\
  \label{eq:demds} 0&=&\widetilde R_{ss}(\theta', q) - R_s(\theta, q) \ \ \ \ \ (q=1,\ldots,Q)
\end{eqnarray}
where we have defined the matrix
\begin{equation}
  \widetilde R_{ss}(\theta, q ) 
  =
  \frac 1 {n_q} \sum_{k\in\dom_q} E\left( s(k)s(k)\adj  |\ y(k), \theta  \right)  
\end{equation}
and its weighted average over all domains
 \begin{equation}
   \widetilde R_{ss}(\theta)  
   =
   \sum_{q=1}^Q \frac{n_q}n \ \widetilde R_{ss}(\theta, q) 
   .
\end{equation}
The same definitions hold for $\widetilde R_{sy}(q,\theta )$ (resp.
$\widetilde R_{yy}(q,\theta )$) as an averaged conditional expectation of
$s(k)y(k)\adj$ (resp. $y(k)y(k)\adj$) and $\widetilde R_{sy}(\theta)$
(resp.  $\widetilde R_{yy}(\theta )$) as its weighted average over
spectral domains.

Equations~(\ref{eq:demda}-\ref{eq:demds}) are readily solved for
unconstrained $A$, $R_n$ and $R_s(q)$.  Recall however that our model
involves \emph{diagonal} covariance matrices so that the actual
parameter set is $(A, C_j(q), \sigma_d^2)$).  This constraint,
however, preserves the simplicity of the solution of the M~step since
it suffices to use the diagonal parts of the solutions
of~(\ref{eq:demda}-\ref{eq:demds}).  Thus, the M~step boils down to
\begin{eqnarray}
  \label{eq:upda} A          &=& \widetilde R_{ys}^{ } (\theta')  \widetilde R_{ss} (\theta')  \inv \\
  \label{eq:updn} \sigma_i^2 &=& \left[ \widetilde R_{yy} (\theta')  - \widetilde R_{ys}(\theta') \widetilde R_{ss}(\theta')\inv \widetilde R_{sy}(\theta')\right]_{ii}\\
  \label{eq:upds} P_i(q)     &=& [\widetilde R_{ss}  (\theta',q )]_{ii}   
\end{eqnarray}

The E-step of the algorithm essentially consists in computing the
conditional covariance matrices $\widetilde R_{\times\times}(q)$.  In this
step again, the linearity and the Gaussianity of the model, together
with the domain approximation, again provides us with significant
computational savings.
Indeed, matrices $C$ and $W$ defined at eqs.~(\ref{eq:defC})
and~(\ref{eq:defW}) are actually constant over each spectral domain so
that the E-step is implemented by the following computations which
directly stem from~(\ref{eq:defC}-\ref{eq:defW}) and from
eqs.(\ref{eq:esscy}-\ref{eq:eyycy}) :
\begin{eqnarray}
  \label{eq:defCq} C(q) &=& (A\adj R_n\inv A + R_s (q) \inv )\inv                 \\
  \label{eq:defWq} W(q) &=& (A\adj R_n\inv A + R_s (q) \inv )\inv  A\adj R_n\inv  \\
        \widetilde R_{ss}(q) &=& W(q) \hat R_y(q) W(q)\adj +  C(q)  \\
        \widetilde R_{sy}(q) &=& W(q) \hat R_y(q)  
\end{eqnarray}

From this, one easily reaches the EM algorithm as described at
algorithm~\ref{algo:1}.  The description of this procedure is
completed by specifying the initialization, the rescaling of the
parameters and the stopping rule, as briefly discussed next.

\mypar{Some comments on EM implementation}

Rescaling is required because, as noted above, the model is not
completely identifiable: the spectral density matrices $R_Y$ are
unaffected by the exchange of a scalar factor between each column of
$A$ and each component's power spectrum.  We have found that this
inherent indetermination must be fixed in order for EM to converge.
Our strategy is, after each EM~step, to fix the norm of each column of
$A$ to unity and to adjust the corresponding power spectra
accordingly.  This is an arbitrary choice which happens to work well
in practice.

The algorithm is initialized with the following parameters. We take
$R_n$ to be $\diag(\widehat R_y)$ where $\widehat R_y= \sum_q \frac
{n_q}n \widehat R_y(q)$.  This is a gross overestimation since it amounts
to assume no signal and only noise.  The initial value of $A$ is
obtained by using the $N_c$ dominant eigen-vectors of $\widehat R_y$
as the $N_c$ columns of $A$.  Again, this is nothing like any real
\emph{estimate} of $A$, but rather a vague guess in `the right
direction'.  Finally, the spectra $P_i(q)$ are taken as to be the
diagonal entries of $A\adj\widehat R_y(q)A$ which would be a correct
estimate in the noise free case if $A$ itself was.  This ad hoc
initialization procedure seems satisfactory.  Note that it is a common
rule of thumb to initialize EM with overestimated noise power.

Regarding the stopping rule, recall (from sec.~\ref{sec:nlin-optim})
that the EM algorithm is only used `halfway' to the maximum of the
likelihood and maximization is completed by a quasi-Newton technique
For this reason, there is little point in devising a sophisticated
stopping strategy: in practice, the algorithm is run for a
pre-specified number of steps (based on a few preliminary experiments
with the data).



\label{lastpage}

\end{document}